\documentclass[journal]{IEEEtran}
\usepackage{color}

\usepackage{cite}

\ifCLASSINFOpdf
   \usepackage[pdftex]{graphicx}
\else
\fi
\usepackage{amsmath}
\usepackage{algorithmic}
\begin{document}
%
\title{Fiber radio frequency transfer using bidirectional frequency division multiplexing dissemination}
%
%
%

\author{Qi Li,
        Liang Hu,~\IEEEmembership{Member,~IEEE,}
        Jinbo Zhang,
        Jianping Chen,~\IEEEmembership{Member,~IEEE,}
        and Guiling Wu,~\IEEEmembership{Member,~IEEE}
        
\thanks{Manuscript received xxx xxx, xxx; revised xxx xxx, xxx. This work was supported by the National Natural Science Foundation of China (NSFC) (61627817, 61905143). \textit{(Corresponding
author: Liang Hu.)}}
\thanks{The authors are with the State Key Laboratory of Advanced Optical Communication Systems and Networks, Shanghai Institute for Advanced Communication and Data Science, Shanghai Key Laboratory of Navigation and Location-Based Services, Department of Electronic Engineering, Shanghai Jiao Tong University, Shanghai 200240, China (email: liang.hu@sjtu.edu.cn.)}
}

%
%

\markboth{IEEE PHOTONICS TECHNOLOGY LETTERS,~Vol.~XX, No.~XX, XX~XXX}%
{Shell \MakeLowercase{\textit{et al.}}: Bare Demo of IEEEtran.cls for IEEE Journals}

\maketitle

\begin{abstract}
We report on the realization of a novel fiber-optic radio frequency (RF) transfer scheme with the bidirectional frequency division multiplexing (FDM) dissemination technique. Here, the proper bidirectional frequency map used in the forward and backward directions for suppressing the backscattering noise and ensuring the symmetry of the bidirectional transfer RF signals within one telecommunication channel. We experimentally demonstrated a 0.9 GHz signal transfer over a 120 km optical link with the relative frequency stabilities of $2.2 \times 10^{-14}$ at 1 s and $4.6 \times 10^{-17}$ at 20,000 s. The implementation of phase noise compensation at the remote site has the capability to perform RF transfer over a branching fiber network with the proposed technique as needed by large-scale scientific experiments.
\end{abstract}


%
\IEEEpeerreviewmaketitle

\section{Introduction}
%
%
%
%
\IEEEPARstart{U}{ltra-stable} radio frequency (RF) dissemination plays a significant role in numerous areas, such as radio astronomy, accurate navigation and basic physical measurement \cite{1,2,3}. Compared with satellite links, fiber-optic links have the advantages of low attenuation, wide bandwidth and immunity to electromagnetic interference, etc \cite{4}, which enable ultra-long-haul and high-precision frequency dissemination \cite{PRL,THU,JP,7,11,13,14,AUS}. Unfortunately, the temperature variations and mechanical perturbations along the fiber links will change the propagation delay of the fiber links and hence introduce additional phase noise into the transferred RF signal \cite{sliwczynski2012bidirectional}. In order to suppress this part of phase noise, various fiber-optic RF transfer schemes based on active \cite{PRL,7,THU,JP} and passive phase noise compensation schemes \cite{11,13,14,AUS} have been proposed by several research groups over the last two decades.

The limits of the short-term stability for the single-fiber bidirectional fiber-optic RF transfer system are mainly decided by system's signal-to-noise ratio (SNR) \cite{sliwczynski2012bidirectional}. The deterioration of the system SNR is generally caused by the backscattering noise such as Fresnel reflection, Rayleigh and Brillouin scattering for single-fiber bidirectional dissemination systems. The bidirectional wavelength division multiplexing (WDM) technique at the local and remote sites has been widely adopted to largely suppress the backscattering noise at the cost of breaking the wavelength symmetry in both directions \cite{13,14}. However, the telecommunication channels in the metropolitan network are scarce resources, this WDM method may occupy an additional telecommunication channel. Alternatively, the bidirectional frequency division multiplexing (FDM) technique has drawn widely attention for the fiber-optic RF transfer application \cite{lopez,7,11}, in which only one telecommunication channel (e.g.100 GHz) is used for bidirectional transmission. Lopez \textit{et al.} have proposed an active phase noise compensation scheme based on the variable optical delay lines (VODLs), which adopted the different frequencies and same wavelength in the bidirectional directions to largely suppress the backscattering noise \cite{lopez}. However, the used VODLs are cumbersome and have the limited compensation range. Huang \textit{et al.} proposed a passive compensation scheme by using frequency dividing and filtering \cite{11}. Although this scheme can effectively eliminate crosstalk and partially suppress the backscattering noise with the same wavelength at both sides, the two modulated RF signals used at the local site have the multiple relationship, such as 4 times or 5 times. Zhang \textit{et al.} have proposed an active electrical compensation scheme using a carrier suppressed double-sideband (CSDSB) signal \cite{7}. Although the CSDSB signal used at the local site can efficiently suppress the backscattering noise, the modulator has to be biased at the $V_{\pi}$ point, which is easily affected by the temperature fluctuations. In this configuration, the bidirectional RF signals have a factor of 2 difference and the RF signal with the lower power will be obtained through mutual-beating between double-sidebands, causing the lower SNR and limiting the system performance, in particular the short-term stability, for long-haul fiber link dissemination. We can see that the existing bidirectional FDM dissemination based fiber-optic RF transfer schemes have the multiple relationship between the bidirectional RF signals, resulting in that the performance of system in terms of the relative stability will be determined by the lowest frequency signal among the bidirectional RF signals \cite{11}. One intriguing question is how to distribute a reference RF to the user by using bidirectional FDM dissemination with the improved performance in a cost-effective and robust way as optical frequency transfer does \cite{Ma:94, Hu:20}.

In this article, we propose a novel active phase noise compensation scheme with the bidirectional FDM technique. The effect of the backscattering noise is effectively suppressed by properly allocating the forward and backward RF signals. The bidirectional optical signals transmitted over the same fiber within one telecommunication channel guarantee the bidirectional propagation symmetry. Moreover, the active compensation devices placed at the remote site can greatly simplify the configuration of the local site and have the capability to expand into the tree topology for multiple-point stable RF optical transmission \cite{9, hu2021all}. We experimentally demonstrate the scheme by transferring a 0.9 GHz RF signal via 120 km single-mode fibers (SMFs), achieving the relative frequency stabilities of approximately $2.2 \times 10^{-14}$ at 1s and $4.6 \times 10^{-17}$ at 20,000s.

\section{Principle}
\begin{figure}[t!]
\begin{center}
\includegraphics[width=0.92\linewidth]{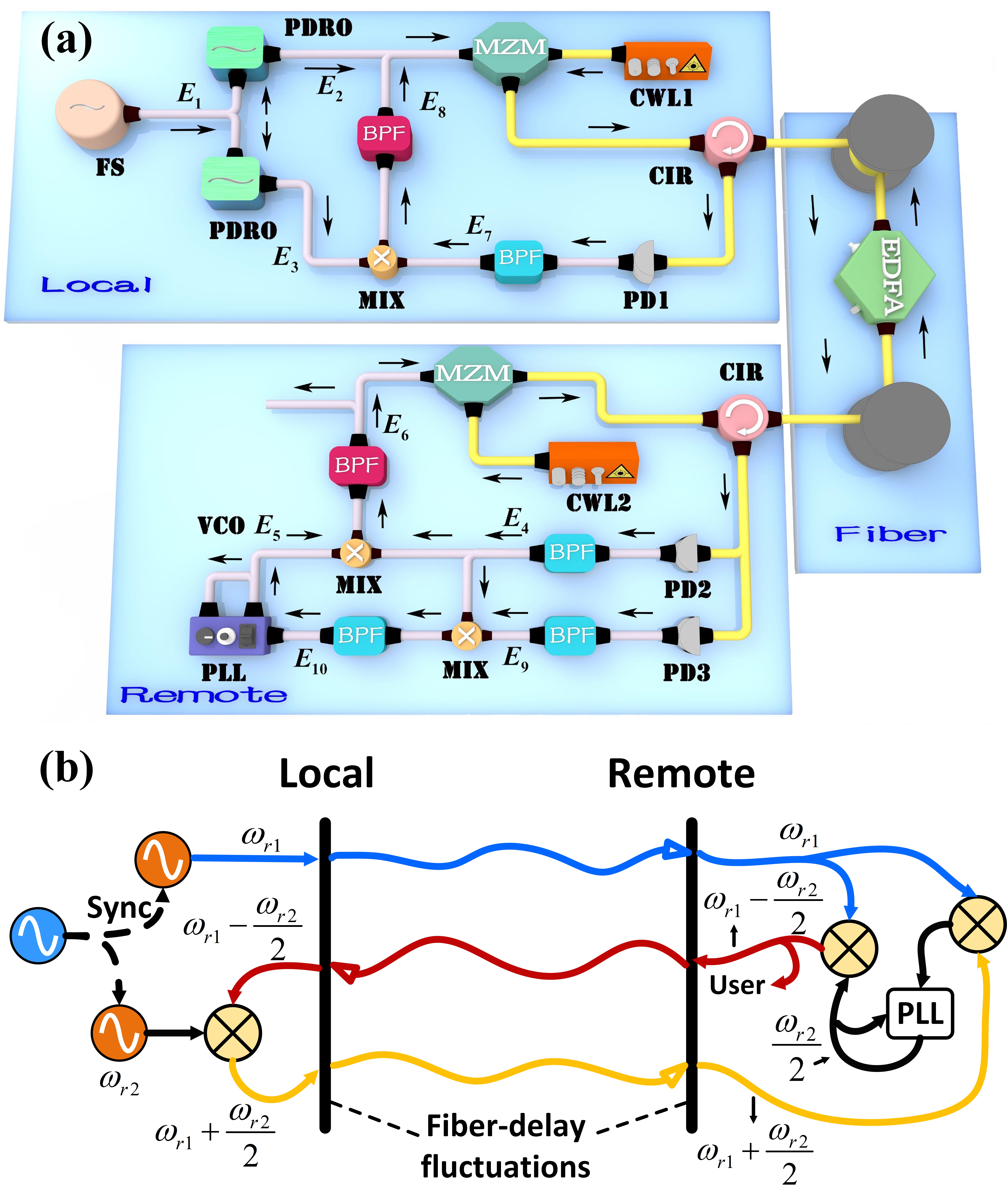}
\end{center}
\caption{(a) Schematic diagram of the proposed fiber-optic RF transfer scheme based on the bidirectional frequency division multiplexing (FDM) transmission. FS: frequency standard, PD: photodetector, MZM: Mach-Zehnder modulator, CIR: optical circulator, VCO: voltage-controlled oscillator, MIX: frequency mixer, PDRO: phase-locked dielectric resonant oscillator, BPF: band-pass filter, EDFA: erbium-doped fiber amplifier, CWL: continuous wave laser, PLL: phase-lock loop. (b) Diagram view of the bidirectional FDM technique to suppress the backscatteringnoise.}
\end{figure}

Figure 1 illustrates a schematic diagram of our proposed stable RF transfer system using bidirectional FDM dissemination. The local site (LS) and remote site (RS) are connected via a fiber-optic link. At the LS, the frequency standard (FS) can be denoted as a cosine function without considering its amplitude \cite{14}, 
\begin{equation}
E_1\propto\cos(\omega_{r0}t+\varphi_{r0}), 
\end{equation}
where $\omega_{r0}$ and $\varphi_{r0}$ are its angular frequency and initial phase, respectively. To improve the stability of the RF transfer system,  the frequency of $E_1$ is multiplied to $E_2$ by a phase-locked dielectric resonant oscillator (PDRO). At the same time, $E_1$ is multiplied to $E_3$ as an assistant signal for achieving the bidirectional FDM dissemination with another PDRO. $E_2$ and $E_3$ can be expressed as, 
\begin{equation}
{E_2} \propto \cos ({\omega _{r1}}t + {\varphi _{r1}}),\label{1}
\end{equation}
\begin{equation}
{E_3} \propto \cos ({\omega _{r2}}t + {\varphi _{r2}}).\label{2}
\end{equation}

$E_2$ is modulated onto the continuous wave laser (CWL1) by a Mach-Zehnder intensity modulator (MZM) biased at the quadrature point and subsequently propagated through the optical fiber with the length of $L$ to the photodetector (PD2) at the RS. The recovered signal has a form of,
\begin{equation}
{E_4} \propto \cos [{\omega _{r1}}(t - {\tau _{LR}}(t)) + {\varphi _{r1}}],\label{3}
\end{equation}
where $\tau_{LR}(t)$ represents the fiber-delay variations induced by the environmental perturbations propagating from the LS to the RS. The detected RF signal is split two parts. One part is mixed with the $E_5$ signal produced by a voltage-controlled oscillator (VCO) with $E_5 \propto cos(\omega_{vco}t+\varphi_c)$, where $\omega_{vco}$ and $\varphi_c$ respectively, are the angular frequency and initial phase of the VCO. To satisfy the bidirectional FDM transmission requirements, we set $\omega_{r2}=2\omega_{vco}$. By mixing $E_4$ with $E_5$, the lower side-band signal is selected by a band-pass filter and can be expressed as,
\begin{equation}
{E_6} \propto \cos [{\omega _{r1}}(t - {\tau _{LR}}(t)) - \frac{1}{2}{\omega _{r2}}t - {\varphi _c} + {\varphi _{r1}}].\label{4}
\end{equation}

Subsequently, $E_6$ is divided into two branches. One part is used for the user and the other part is modulated onto the CWL2 by another MZM biased at the $V_{\pi/2}$ point and transmitted to the LS via the same SMFs. To get rid of the interference from the backscattered optical signal carrier in the bidirectional fiber-optic link, we can set the wavelengths of the CWL1 and CWL2 with the slight difference. The RF signal recovered by the PD1 at the LS can be depicted as,
\begin{equation}
\begin{split}
{E_7} \propto \cos [{\omega _{r1}}(t - {\tau _{RL}}(t) - {\tau _{LR}}(t)) 
\\- \frac{1}{2}{\omega _{r2}}(t - {\tau _{RL}}(t)) - {\varphi _c} + {\varphi _{r1}}],\label{5}
\end{split}
\end{equation}
where $\tau_{RL} (t)$ represents the fiber-delay variations from the RS to the LS. The upper side-band signal obtained by mixing $E_3$ and $E_7$ can be expressed as,
\begin{equation}
\begin{split}
{E_8} \propto \cos [{\omega _{r1}}(t - {\tau _{RL}}(t) - {\tau _{LR}}(t)){\rm{}} 
\\- \frac{1}{2}{\omega _{r2}}(t - {\tau _{RL}}(t)) + {\omega _{r2}}t - {\varphi _c} + {\varphi _{r1}} + {\varphi _{r2}}].\label{6}
\end{split}
\end{equation}

Then $E_2$ and $E_8$ are simultaneously modulated onto the CWL1 by the MZM at the LS and transmitted to the RS via the fiber link, the signal recovered by the PD3 can be denoted as,
\begin{equation}
\begin{split}
{E_9} \propto \cos [{\omega _{r1}}(t - {\tau _{RL}}(t) - 2{\tau _{LR}}(t)) + {\omega _{r2}}(t - {\tau _{LR}}(t))\\
{\rm{}} - \frac{1}{2}{\omega _{r2}}(t - {\tau _{RL}}(t) - {\tau _{LR}}(t)) - {\varphi _c} + {\varphi _{r1}} + {\varphi _{r2}}].\label{7}
\end{split}
\end{equation}

By mixing $E_4$ with $E_9$, the lower side-band signal is obtained and can be given by,
\begin{equation}
\begin{split}
{E_{10}} \propto \cos [\frac{1}{2}\omega _{r2}t - ({\omega _{r1}} - \frac{1}{2}{\omega _{r2}}){\tau _{RL}}(t)\\
{\rm{}} - ({\omega _{r1}} + \frac{1}{2}{\omega _{r2}}){\tau _{LR}}(t) - {\varphi _c} + {\varphi _{r2}}]
\end{split}.\label{8}
\end{equation}

In Eq. \ref{8}, the DC error signal carrying the fiber link phase noise is obtained by mixing $E_{5}$ and $E_{10}$, resulting in ,
\begin{equation}
{V_{error}} \propto ({\omega _{r1}} - \frac{1}{2}{\omega _{r2}}){\tau _{RL}}(t){\rm{ + }}({\omega _{r1}} + \frac{1}{2}{\omega _{r2}}){\tau _{LR}}(t) + 2{\varphi _c} - {\varphi _{r2}}.\label{9}
\end{equation}

Assuming the fiber-delay variations introduced by the environmental perturbations slower than the triple transmission time of the fiber link, we can have the relationship of $\tau_{LR}(t)=\tau_{RL}(t)=\tau(t)$. As the first and second terms indicated in the right side of Eq. \ref{9}, the phase noise of the fiber-optic link caused by forward and backward transfer is centrally symmetric about the angular frequency $\omega_{r1}$ as indicated by Fig. 1(b). Equation \ref{9} can be further rewritten as $V_{error}\propto 2\omega_{r1}\tau(t) +2\varphi_c-\varphi_{r2}.$ When the phase-locked loop (PLL) is active, the fiber link is stabilized by tuning the frequency of the VCO, i.e., $V_{err} \rightarrow 0$. We obtain the relationship of,
\begin{equation}
{\varphi _c} = \frac{{{\varphi _{r2}}}}{2} - {\omega _{r1}}\tau (t).\label{10}
\end{equation}

By substituting the Eq. \ref{10} into Eq. \ref{4}, we can find that a stable RF signal can be recovered at the RS, which is independent of the fiber-optic induced phase fluctuations. The principle of the proposed scheme to suppress the backscattering noise can be seen in Fig. 1(b). As the frequencies of the RF signals transmitted through the fiber-optic link are different on the bidirectional directions, and these RF signals are symmetric about the center of $\omega_{r1}$, the backscattering noise can be suppressed by electrical filters.

\section{Experimental apparatus and results}

The experiment setup based on the proposed scheme is shown in Fig. 1. At the LS, the frequency standard (Rigol Inc., DSG 830) with $\omega_{r0}=100\ \rm{MHz}$ is multiplied to $\omega_{r1}=1\ \rm{GHz}$ and $\omega_{r2}=200\ \rm{MHz}$ by the PDROs. The LS and the RS are connected by 120 km SMFs (consisting of 10, 30, 50, 30 km single-mode fiber spools), placed in a normal laboratory with the peak-to-peak temperature fluctuations about 3 $^{\circ}$C \cite{19, spy}. To suppress the impact of the chromatic dispersion, the proper dispersion compensating fibers (DCFs) are used \cite{19, spy}. We set the wavelengths of the CWLs with the 0.4 nm difference at the both sites around 1550 nm. The wavelength difference is determined by the minimum adjustment step size of the lasers used in our experiment. Each transmitted RF signal is carried on the corresponding optical carrier by a LiNbO3 MZM (iXblue Inc., MXAN-LN-10) biased at the $V_{\pi/2}$ point and detected by the PD with 2 GHz bandwidth at each site. With this configuration, we have $\omega_{r1}=1\ \rm{GHz}$, $\omega_{r1}-\omega_{r2}/2=0.9\ \rm{GHz}$ and $\omega_{r1}+\omega_{r2}/2=1.1\ \rm{GHz}$. The PID controller (Newport Inc., LB1005) as the PLL control unit is used and we set the corner frequency of approximately 200 Hz. Moreover, a bidirectional erbium-doped fiber amplifier (Bi-EDFA) \cite{lopez} is employed between the 40 km and 80 km fiber spools, to boost the fading bidirectional optical signals. The optical gain of Bi-EDFA at each direction is less than 20 dB to suppress the stimulated Brillouin scattering and amplified spontaneous emission (ASE) noise. To evaluate the performance of the proposed scheme, the standard signal at the LS and the duplicated 0.9 GHz signal at RS are converted to 10 MHz by a dual-mixer time difference method \cite{JP}. The outputs are measured by a phase noise measurement instrument (Symmetricom Inc., TSC5120A).

\begin{figure}[htbp]
\centering\includegraphics[width=0.92\linewidth]{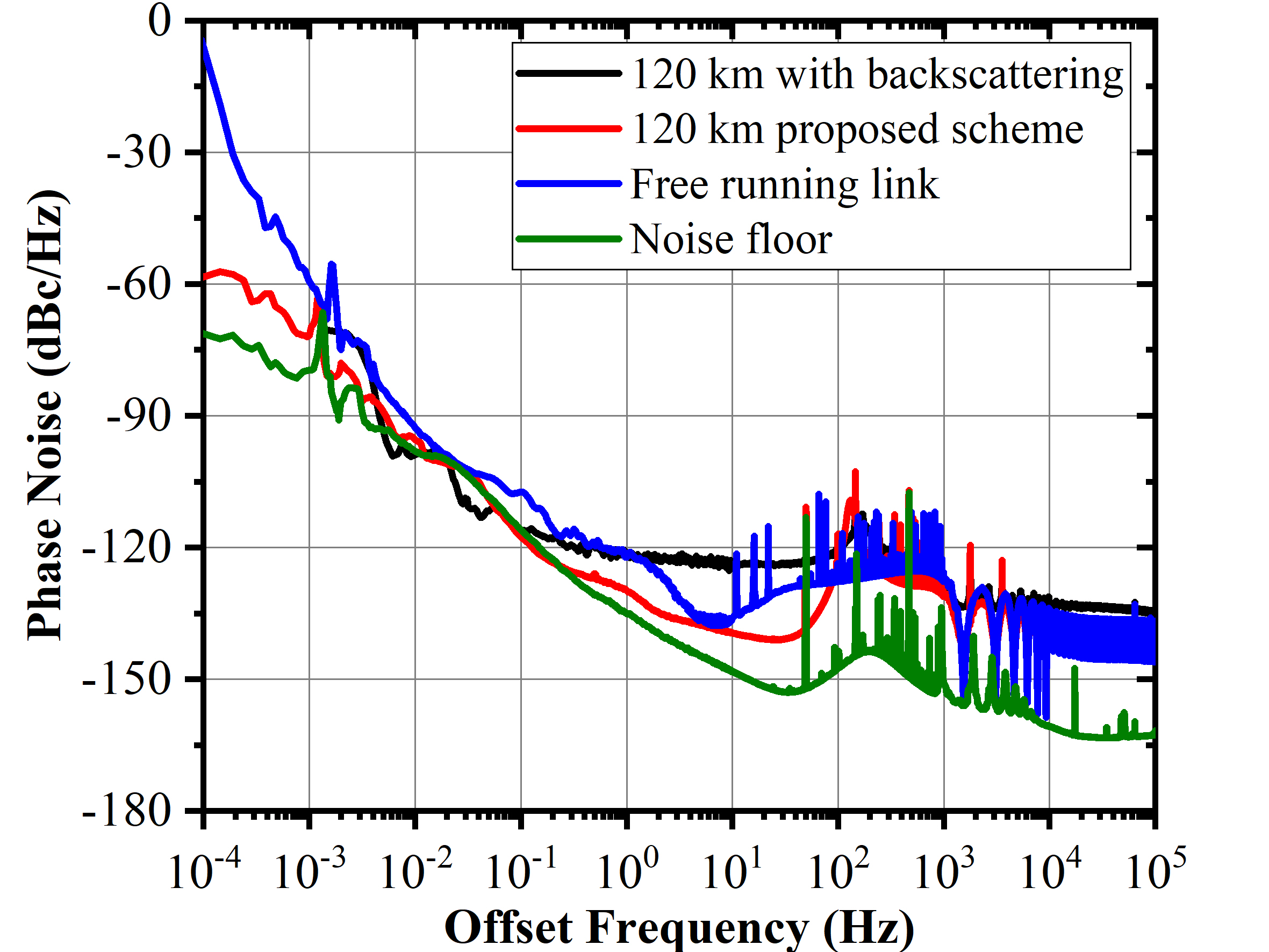}
\caption{Measured phase noise PSDs for different system configurations: the 120 km free running link (blue curve), the 120 km proposed scheme (red curve), the back-to-back system noise floor (green curve) and the 120 km compensated link with the backscattering noise (black curve).}
\end{figure}

The measured phase noise power spectral densities (PSDs) of the system under different configurations are illustrated in Fig. 2. As a comparison, a RF transfer scheme with the backscattering noise by adopting the same 1 GHz signal for the bidirectional transmission is measured. The configuration is similar to Ref \cite{10}. The back-to-back system noise floor is also measured by connecting two sites of the proposed scheme with a 1 m fiber link. The 120 km free running link is evaluated by transferring a 0.9 GHz signal to the RS without round-trip phase correction. As depicted in Fig. 2, the phase noise of the 120 km free running link is obviously higher than the 120 km proposed scheme (about -117 dBc at 0.1 Hz, -129 dBc at 1 Hz) at the offset frequency of less than 1 Hz, due to the fiber delay fluctuations caused by the temperature variations. In the range of $0.002-0.1\ \rm{Hz}$, the 120 km proposed compensated link is almost consistent with the system noise floor, which indicates that the phase noise in this low frequency range induced by the fiber-optic link can be effectively suppressed. Compared with the 120 km compensated scheme with the backscattering noise (about -116 dBc at 0.1 Hz, -121 dBc at 1 Hz), the phase noise of the 120 km proposed scheme is significantly lower within $0.1-100\ \rm{Hz}$, demonstrating that the backscattering noise induced by the optical reflections including connectors and the Rayleigh backscattering has a significant effect on the system performance \cite{19}. All the measurements have the sharp bump in the range of $0.001-0.003\ \rm{Hz}$, which are mainly affected by the outside loop phase noise induced by the temperature fluctuations. The phase noise of all compensated configurations has a bump observed at about 200 Hz. This part noise is mainly determined by the PLL parameters, the phase noise of the RF signal and the propagation delay from the fiber \cite{19}. At the high frequency range ($> 10^3$ Hz), the ripples with the same period can be observed in different transfer configurations under the 120 km fiber link, which are mainly caused by delayed self-interferometry \cite{21}.

\begin{figure}[htbp]
\centering\includegraphics[width=0.92\linewidth]{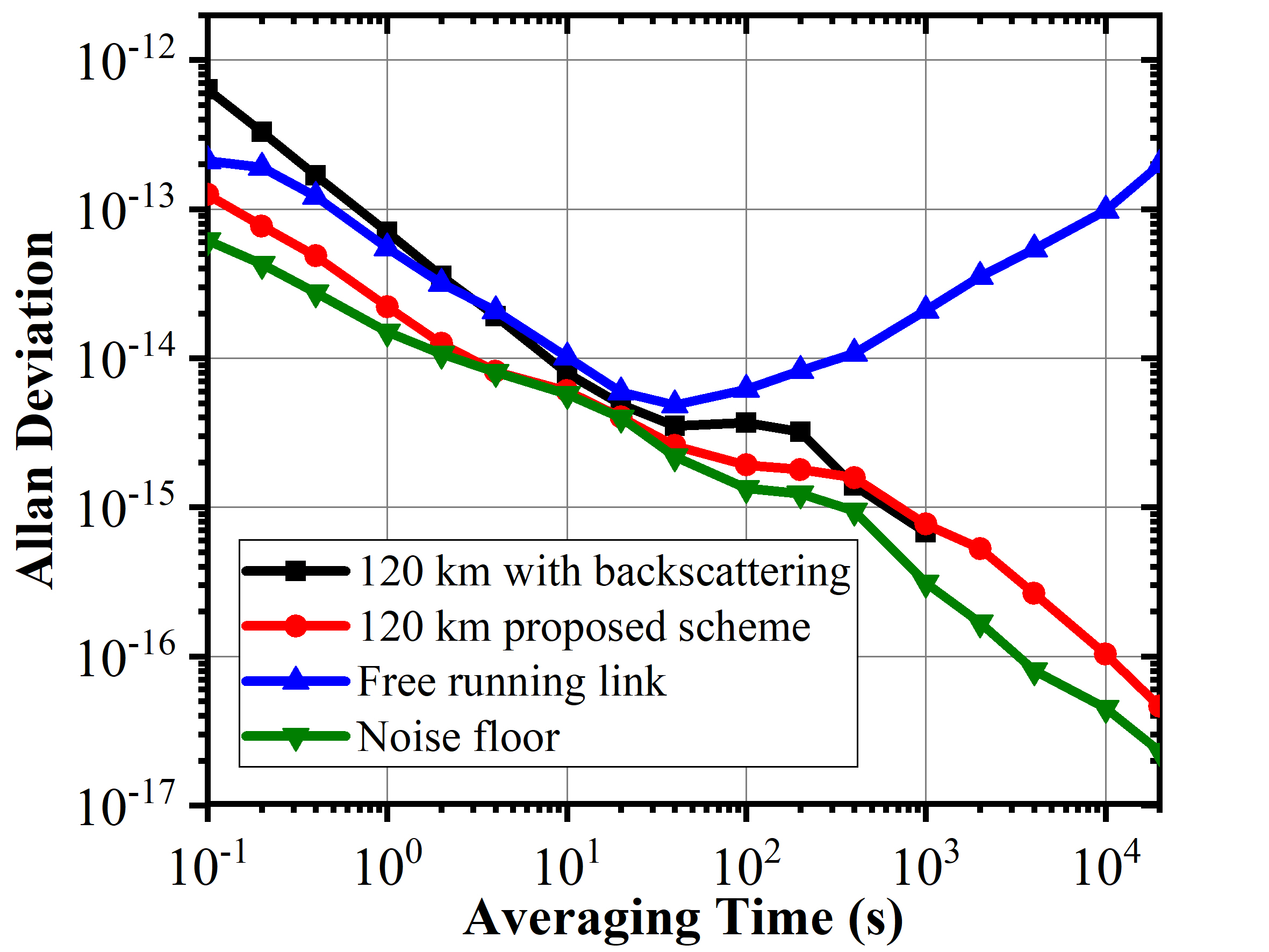}
\caption{Measured Allan deviations in different system configurations: the 120 km free running link (blue curve), the 120 km proposed scheme (red curve), the back-to-back system noise floor (green curve) and the 120 km compensated link with the backscattering noise (black curve).}
\end{figure}

Figure 3 shows the measured Allan deviations for different system configurations with a 5 Hz measurement bandwidth. It is seen that, the long-term stability of $4.6\times10^{-17}$ at the averaging time of 20,000 s is achieved for the 120 km proposed scheme, which is improved more than four orders of magnitude in contrast with the 120 km free running link. Compared with the 120 km compensated scheme with the backscattering noise, the short-term of the 120 km proposed scheme is improved from $7.0\times10^{-14}$ to $2.2\times10^{-14}$ at the averaging time of 1 s. Moreover, this value is close to the back-to-back system noise floor, illustrating the fact that the backscattering noise in the fiber link is largely suppressed by the proposed scheme. Whereas, the Allan deviation of the 120 km proposed scheme is slightly larger than the system noise floor, due to the noise of the source signals and RF components, the ASE noise introduced by the EDFA and so on \cite{20,spy}. As the outside loop of the system is easily susceptible to the temperature fluctuations, the Allan deviations of the different compensated systems decrease slowly within $10^2-10^3\ \rm{s}$, which correspond to the sharp bump at the range of $0.001-0.003\ \rm{Hz}$ in Fig. 2 \cite{8}.

\section{Conclusion}
In conclusion, we proposed and demonstrated a novel fiber-optic RF transfer scheme with the bidirectional FDM transmission technique. In the bidirectional fiber-optic link, the frequencies of the RF signals transmitted on both directions are different and have a proper relationship, which can efficiently suppress the effect of the backscattering noise and settle out the issues that the system performance is limited to the lowest transfer frequency signal in the conventional FDM transmission schemes with the multiple relationship for the bidirectional frequency signals. We experimentally demonstrated a 0.9 GHz signal transfer based on the proposed scheme over a 120 km fiber-optic link. The stable 0.9 GHz signal transferred to the remote site has the relative frequency stabilities of less than $2.2\times10^{-14}$ at 1 s and $4.62\times10^{-17}$ at 20,000 s. The phase noise compensation devices placed at the remote site, which are compatible with the tree topology, provide a guidance for developing multiple-point RF transfer as required by many scientific applications.


%



\ifCLASSOPTIONcaptionsoff
  \newpage
\fi



\bibliographystyle{IEEEtran}



%

%




\end{document}